\documentclass{article}

\usepackage{setspace}
\usepackage{amssymb}
\usepackage{amsmath}
\usepackage{amsfonts}
\usepackage{epstopdf}
\usepackage{placeins}
\usepackage{color}
\usepackage{float}
\usepackage{bbm}
\usepackage{tikz}
\usetikzlibrary{arrows.meta}% arrow tip library
\usepackage{geometry}
\begin{document}

\title {On Universal Eigenvalues of Casimir Operator}
\author{M.Y.Avetisyan }
\date{}

 \maketitle
 
 %\vspace{-6.0cm}

%\begin{center}

%\end{center}

%\vspace{4.2cm}

\begin{center}
 {\small {\it Yerevan Physics Institute, Yerevan, Armenia}}

\end{center}

\vspace{1cm}

{\small  {\bf Abstract.} Motivated by the universal knot polynomials in the gauge Chern-Simons theory, we show that the values of the second Casimir operator on an arbitrary power of Cartan product of $X_2$ and adjoint 
representations of simple Lie algebras can be represented in a universal form. We show that it complies with 
 $N\longrightarrow -N$ duality  of the same operator for $SO(2n)$ and $Sp(2n)$  algebras (the part of $N\leftrightarrow-N$  duality of gauge $SO(2n)$ and $Sp(2n)$   theories). We discuss the phenomena of non-zero
  universal values of Casimir operator on zero representations. }

\tableofcontents

\newpage

\section{Introduction}

The universal formulae for simple Lie algebras were first derived by P. Vogel in his Universal Lie Algebra \cite{V0,V}. The main aim was to derive the most general weight system for Vassiliev's finite knot invariants. This program met difficulties, however, as a byproduct there appeared the uniform parameterization of simple Lie algebras by the values of Casimir operators on three representations, appearing in decomposition of the symmetric square of the adjoint representations:

\begin{eqnarray}\label{sad}
S^2 \mathfrak{g}=1+Y_2(\alpha)+Y_2(\beta)+Y_2(\gamma)
\end{eqnarray}

One denotes the value of the second Casimir operator on the adjoint representation $\mathfrak{g}$ as $2t$, and parameterizes the values of the same operator on representations in (\ref{sad}) as $4t-2\alpha, 4t-2\beta, 4t-2\gamma$ correspondingly (hence notation of representations in (\ref{sad})). It appears that $\alpha+\beta+\gamma=t$. The values of the parameters for all simple Lie algebras are given in the table 1, and in the table \ref{tab:V2} in another form. According to the definitions, the entire theory is invariant with respect to rescaling of the parameters (which corresponds to rescaling of invariant scalar product in algebra), and with respect to the permutation of the universal (=Vogel's) parameters  $\alpha, \beta, \gamma$. So, effectively they belong to a projective plane, which is factorized w.r.t. its homogeneous coordinates, and is called Vogel's plane. 
\begin{table}[ht]
\caption{Vogel's parameters for simple Lie algebras}     \label{tab:V1}

\begin{tabular}{|c|c|c|c|c|c|}
\hline
Root system & Lie algebra  & $\alpha$ & $\beta$ & $\gamma$  & $t$\\   
\hline    
$A_n$ &  $\mathfrak {sl}_{n+1}$     & $-2$ & 2 & $(n+1) $ & $n+1$\\
$B_n$ &   $\mathfrak {so}_{2n+1}$    & $-2$ & 4& $2n-3 $ & $2n-1$\\
$C_n$ & $ \mathfrak {sp}_{2n}$    & $-2$ & 1 & $n+2 $ & $n+1$\\
$D_n$ &   $\mathfrak {so}_{2n}$    & $-2$ & 4 & $2n-4$ & $2n-2$\\
$G_2$ &  $\mathfrak {g}_{2}  $    & $-2$ & $10/3 $& $8/3$ & $4$ \\
$F_4$ & $\mathfrak {f}_{4}  $    & $-2$ & $ 5$& $ 6$ & $9$\\
$E_6$ &  $\mathfrak {e}_{6}  $    & $-2$ & $ 6$& $ 8$ & $12$\\
$E_7$ & $\mathfrak {e}_{7}  $    & $-2$ & $ 8$& $ 12$ & $18$ \\
$E_8$ & $\mathfrak {e}_{8}  $    & $-2$ & $ 12$& $20$ & $30$\\
\hline  
\end{tabular}
\end{table}

\begin{table}[ht] 
 \caption{Vogel's parameters for simple Lie algebras: lines}
\begin{tabular}{|r|r|r|r|r|r|} 
\hline Algebra/Parameters & $\alpha$ &$\beta$  &$\gamma$  & $t$ & Line \\ 
\hline  $\mathfrak {sl}_{N}$  & -2 & 2 & $N$ & $N$ & $\alpha+\beta=0$ \\ 
\hline $\mathfrak {so}_{N}$ & -2  & 4 & $N-4$ & $N-2$ & $ 2\alpha+\beta=0$ \\ 
\hline  $ \mathfrak {sp}_{N}$ & -2  & 1 & $N/2+2$ & $N/2+1$ & $ \alpha +2\beta=0$ \\ 
\hline $Exc(n)$ & $-2$ & $2n+4$  & $n+4$ & $3n+6$ & $\gamma=2(\alpha+\beta)$\\ 
\hline 
\end{tabular}

{For the exceptional 
line $n=-2/3,0,1,2,4,8$ for $\mathfrak {g}_{2}, \mathfrak {so}_{8}, \mathfrak{f}_{4}, \mathfrak{e}_{6}, \mathfrak {e}_{7},\mathfrak {e}_{8} $, 
  respectively.} \label{tab:V2}
\end{table}

\newpage

As an example of application of this parametrization universal formulae
\cite{V0,LM1} for dimensions of representations from (\ref{sad}) are presented
below:

\begin{eqnarray}
\text{dim}\, \mathfrak{g} &=&\frac{(2t-\alpha)(2t-\beta)(2t-\gamma)}{\alpha\beta\gamma} \\
 \label{Y}
 \text{dim} \, Y_2(\alpha)&=& \frac{\left(  2t  - 3\alpha \right) \,\left( \beta - 2t \right) \,\left( \gamma - 2t \right) \,t\,\left( \beta + t \right) \,
      \left( \gamma + t \right) }{\alpha^2\,\left( \alpha - \beta \right) \,\beta\,\left( \alpha - \gamma \right) \,\gamma}
\end{eqnarray}

and  other two (\ref{Y}) representations which are obtained by permutations 
of the parameters. These are typical universal formulae for dimensions: 
ratios of products of linear homogeneous functions of universal parameters.

There are a number of universal formulae for different objects in the theory and
applications of simple Lie algebras. E.g. Vogel \cite{V0} found complete
decomposition of third power of the adjoint representation in terms of  Universal
Lie Algebra, defined by himself, and universal dimension formulae for all
representations involved. Landsberg and Manivel \cite{LM1} present a method which
allows derivation of certain universal dimension formulae for simple Lie algebras
and derive those for Cartan powers of the adjoint, $Y_2(.)$, and their Cartan
products.Sergeev, Veselov and Mkrtchyan derived \cite{MSV} a universal formula
for generating function for the eigenvalues of higher Casimir operators on the
adjoint representation. 

In subsequent works applications to physics were developed, particularly the
universality of the partition function of Chern-Simons theory on a sphere
\cite{MV,M13,KhM16-1}, and its connection with q-dimension of $k\Lambda_0$
representation of affine Kac-Moody algebras \cite{M17} were shown. 

Moreover, the universal
knot polynomials for 2- and 3-strand torus knots \cite{W15,MMM,MM,M16QD} were
calculated. This is a partial realization of initial Vogel's program. 

The main motivation for present paper is the extension of construction of invariant knot polynomials of \cite{MMM} to higher-strand torus knots.
The construction of \cite{MMM} was based on the Rosso-Jones formula \cite{RJ}

\begin{eqnarray}
P^{[m,n]}_R = \frac{ q^{mn\varkappa_R}}{D_R(q)} \sum_Y \sum_Q
q^{-\frac{n}{m}\varkappa_{_Q}}  \varphi_{_Y} (\bar{\sigma}^{[m,n]}) D_{_Q}(q)
\label{RJ2}
\end{eqnarray}

Two main ingredients of this formula are eigenvalues $\varkappa_Q$ of second Casimir operators on representation $Q$ and quantum dimensions
 $D_{_Q}(q)$ of the same representation. $Q$ is one of the irreps, appearing in the decomposition of the $m$-th power of representation $R$ with 
 symmetry of a young diagram $Y$ with $m$ boxes, appearing in the decomposition of powers of the adjoint representation. If one wants to get a 
 universal answer, he has to present these two quantities in the universal form. 
 The other elements of the formula such as the character of the symmetric group 
   $\varphi_{_Y} (.)$ are "universal" in the sense that they do not depend on group, and we shall not consider them here. 

We do not consider the universal form of quantum dimensions,  and in this paper will focus on the first quantity - Casimir eigenvalues. 
We present a universal expression for its eigenvalues on certain irreps described below, 
and appearing in Rosso-Jones formula for the case when $R$ is the adjoint representation.

The antisymmetric square of the adjoint representation of semisimple Lie algebras is known to be decomposed in the following universal 
form:
$$\Lambda^2g=g \oplus X_2$$
First of all, let's suppose that for each algebra the square of the long root is equal to $2$. This corresponds to the set of Vogel's 
parameters with $\alpha=-2$.
Having this normalization in mind, we calculate the eigenvalues of the Casimir operator on arbitrary powers of Cartan product of 
$X_2$ and $g$.
The eigenvalue of the Casimir operator on an irrep with $\lambda$ highest weight is equal to $(\lambda+2\rho, \lambda)$.
The corresponding weights are given in the
Table 3, where the same labeling of Dynkin diagrams as in \cite{LieART} is used.\footnote{Note, that for $G_2$ algebra the $\omega_2$ 
weight corresponds to the long root.} 

\newpage

\begin{table}[h]
   \caption{Highest weights of $X_2$ and $g$ representations}
     \begin{tabular}{|c|c|c|}
     \hline
                &$\lambda_{X_2}$&$\lambda_g$\\
     \hline
               $A_n,n\geq3$&$(2\omega_1+\omega_{n-1})\oplus (\omega_2+2\omega_{N})$&$\omega_1+\omega_N$\\
       \hline
               $B_n,n\geq4$&$\omega_1+\omega_3$&$\omega_2$\\
       \hline
        $C_n,n\geq3$&$2\omega_1+\omega_2$&$2\omega_1$\\
       \hline
              $D_n,n\geq5$&$\omega_1+\omega_3$&$\omega_2$\\
       \hline
        $G_2$&$3\omega_1$&$\omega_2$\\
       \hline
       $F_4$&$\omega_2$&$\omega_1$\\
       \hline
       $E_6$&$\omega_3$&$\omega_6$\\
       \hline
       $E_7$&$\omega_2$&$\omega_1$\\
       \hline
    $E_8$&$\omega_6$&$\omega_7$\\
       \hline

       \end{tabular}
       \label{tab:x2acl}
         \end{table}

\newpage

\section{Universal Casimir Eigenvalues for $X_2$ Representation}
Taking into account the above mentioned data, we carry out the direct
calculations for $X_2$ first.
\section*{$A_N$}
$$\alpha=-2, \beta=2, \gamma=N+1, 
\lambda_1=2\omega_1+\omega_{N-1}, \lambda_2=\omega_2+2\omega_{N},$$
For $\lambda_1=2\omega_1+\omega_{N-1}$ case

$$C=(2\omega_1+\omega_{N-1}, 2\omega_1+\omega_{N-1})+2(\omega_1+\dots+\omega_N,2\omega_1+\omega_{N-1})=$$
$$\frac{6N+6}{N+1}+\frac{2}
{N+1}\left(N(N+1)+(N-1)(N+1)\right)=6+2N+2N-2=4N+4$$
For  $\lambda_2=\omega_2+2\omega_{N}$ irrep the Casimir eigenvalue 
coincides with the one calculated above, so the eigenvalue 
on the direct sum
 of 
these two irreps will be $C$.
\section*{$B_N$}
$$\alpha=-2, \beta=4, \gamma=2N-3, \lambda=\omega_1+\omega_3$$
$$C=(\omega_1+\omega_3, \omega_1+\omega_3)+2(\omega_1+\dots+\omega_N,\omega_1+
\omega_3)$$
$$=F_{11}+F_{31}+F_{13}+F_{33}+2\left((F_{11}+F_{12}+\dots+F_{1N})+
F_{31}+\dots+F_{3N}\right)=$$
$$6+2N-1+9+6(N-3)=2N+14+6N-18=8N-4$$
\newline
Where $F_{i,k}=(\omega_i,\omega_k)$.
\section*{$C_N$}
$$\alpha=-2, \beta=1, \gamma=N+2, \lambda=2\omega_1+\omega_2$$
$$C=(2\omega_1+\omega_2, 2\omega_1+\omega_2)+2(\omega_1+\dots+\omega_N,2\omega_1+
\omega_2)$$
$$=4F_{11}+4F_{12}+F_{22}+F_{33}+2\left(2(F_{11}+F_{12}+\dots+
F_{1N})+F_{21}+\dots+F_{2N}\right)$$
$$=
2+2+1+2(N+1/2+N-1)=5+2N+1+2N-2=4N+4$$
\section*{$D_N$}
$$\alpha=-2, \beta=4, \gamma=2N-4, \lambda=\omega_1+\omega_3$$
$C=(\omega_1+\omega_3, \omega_1+\omega_3)+2(\omega_1+\dots+\omega_N,\omega_1+\omega_3)=
6+2(N-1+6+3N-12)=8N-8$
\section*{$G_2$}
$$\alpha=-2, \beta=10/3, \gamma=8/3, \lambda=3\omega_1$$
$C=(3\omega_1, 3\omega_1)+2(\omega_1+\omega_2,3\omega_1)=9F_{11}+2\left(3F_{11}+3F_{12}\right)=
6+4+6=16$
\section*{$F_4$}
$$\alpha=-2, \beta=5, \gamma=6, \lambda=\omega_2$$
$C=(\omega_2, \omega_2)+2(\omega_1+\omega_2+\omega_3+\omega_4,\omega_2)=
6+30=36$
\section*{$E_6$}
$$\alpha=-2, \beta=6, \gamma=8, \lambda=\omega_3$$
$C=F_{33}+2\sum_{k=1}^6F_{3k}=
6+2(6+15)=48$
\section*{$E_7$}
$$\alpha=-2, \beta=8, \gamma=12, \lambda=\omega_2$$
$C=F_{22}+2\sum_{k=1}^6F_{2k}=
6+2(9+14+10)=72$
\section*{$E_8$}
$$\alpha=-2, \beta=12, \gamma=20, \lambda=\omega_6$$
$C=F_{66}+2\sum_{k=1}^6F_{6k}=
6+2\cdot57=120$

It can be easily noticed, that for each of the algebra the obtained 
value can be expressed as $C=4t=4(\alpha+\beta+\gamma).$
\newline
In the work of M.Cohen and R. de Man (\cite{Cohen})  the Casimir 
eigenvalues on each of the irrep appearing in the decomposition of up to 
4th power of the adjoint representation for the exceptional Lie algebras have 
been computed. So we can check the correspondence between our
formula and their calculated one for the exceptional algebras.
If we scale the Casimir eigenvalue to be equal to $1$ on the adjoint representation, 
as it is done in \cite{Cohen}, our formula will be
$C_c=C/2t=4t/2t=2$, which coincides with the value, presented in that work. Below we shall make similar check for other representations, also.

\newpage

\section{Universal Casimir Eigenvalues on Cartan Product of Powers of $X_2$ and $g$}

Now we turn to the Cartan product case.
The highest weights are now $k\lambda_{X_2}$ and $n\lambda_{g}$ correspondingly.
It is easy to check, that 
\begin{equation*}
 C_{k,n}= C_{k\lambda_{X_2}} + C_{n\lambda_{g}}
 + 2kn(\lambda_{X_2},\lambda_g)
 \end{equation*}

Substituting the highest weights in the expression, written above for the Casimir eigenvalue, 
one obtains the expressions shown in the following table:
\begin{table}[h]
   \caption{Casimir Eigenvalues}
   {\small
     \begin{tabular}{|c|c|c|c|c|}
     \hline
       
                    &$C_{k\lambda_{X_2}}$&$C_{n\lambda_{g}}$&$2kn(\lambda_{X_2},\lambda_g)$&$C_{k,n}$ \\
                                     \hline
               $A_N,N\geq3$&$6k^2+k(4N-2)$&$2n(n+N)$&$6kn$&$6k^2+k(4N-2)+2n(n+N)+6kn$\\
       \hline
               $B_N,N\geq4$&$6k^2+k(8N-10)$&$2n(n+2N-2)$&$6kn$&$6k^2+k(8N-10)+2n(n+2N-2)+6kn$\\
       \hline
        $C_N,N\geq3$&$5k^2+k(4N-1)$&$2n(n+N)$&$6kn$&$5k^2+k(4N-1)+2n(n+N)+6kn$\\
       \hline
              $D_N,N\geq5$&$6k^2+k(8N-14)$&$2n(n+2N-3)$&$6kn$&$6k^2+k(8N-14)+2n(n+2N-3)+6kn$\\
       \hline
        $G_2$&$6k^2+10k$&$2n(n+3)$&$6kn$&$6k^2+10k+2n(n+3)+6kn$\\
       \hline
       $F_4$&$6k^2+30k$&$2n(n+8)$&$6kn$&$6k^2+30k+2n(n+8)+6kn$\\
       \hline
       $E_6$&$6k^2+42k$&$2n(n+11)$&$6kn$&$6k^2+42k+2n(n+11)6kn$\\
       \hline
       $E_7$&$6k^2+66k$&$2n(n+17)$&$6kn$&$6k^2+66k+2n(n+17)+6kn$\\
       \hline
    $E_8$&$6k^2+114k$&$2n(n+29)$&$6kn$&$6k^2+114k+2n(n+29)+6kn$\\
       \hline
Universal Form&$ 3\alpha(k-k^2)+4tk   $&$ \alpha (n-n^2)+2tn  $& $-3\alpha kn$&$\alpha(3k-3k^2+n-n^2-3kn)+t(4k+2n)$\\

       \hline
       \end{tabular}
       }
       \label{tab:x2acl}
         \end{table}     
         \newline

One can easily check, that for each of these cases (except for the $C_N$) the  $C_{k\lambda_{X_2}}$ eigenvalue can be expressed as
 $$C_{k\lambda_{X_2}}=-3\alpha k^2+(4t+3\alpha)k=
3\alpha(k-k^2)+4tk,$$ 
for $C_{n\lambda_{g}}$ 
$$C_{n\lambda_{g}}=-\alpha n^2+2n(t-1)=\alpha (n-n^2)+2tn $$

and $$2kn(\lambda_{X_2},\lambda_g)=-3kn\alpha$$
Notice, that when $k=1$,  $C_{X_2}=4t$ and when $n=1$, $C_{g}=2t$, as expected.
\newline
  Finally, the universal formula for the Casimir eigenvalues on the Cartan powers of $X_2$ and $g$ representations is
  $$C_{k,n}=3\alpha(k-k^2)+4tk+\alpha (n-n^2)+2tn=\alpha(3k-3k^2+n-n^2-3kn)+t(4k+2n).$$
 
\subsection{Conformity Check}

Now we turn to the comparison of our universal expression with the values presented in \cite{Cohen}.
The representations on which the Casimir eigenvalues are to be compared are those defined
with the following highest weights: $2\lambda_{X_2}, \lambda_{X_2}+\lambda_g$ and $\lambda_{X_2}+2\lambda_g$.
So, we calculate $\gamma(H), \gamma(C), \gamma(G)$ (i.e. Casimirs in \cite{Cohen} notation) and compare them with
$C_{2,0}, C_{1,1}, C_{1,2}$, written in the corresponding scaling.
\newline
For the $k=2$ and $n=0$ case our formula in the corresponding scaling gives: 
$$C_{2,0}=\frac{3\alpha(k-k^2)+4tk}{2t}=\frac{-3\alpha+4t}{t}=\frac{6+4t}{t}.$$
For $k=1, n=1$ 
$$C_{1,1}=\frac{6t-3\alpha}{2t}=\frac{3(t-1)}{t}$$
Finally, for $k=1,n=2$
$$C_{1,2}=\frac{-8\alpha+8t}{2t}=\frac{8+4t}{t}$$

In the following table the corresponding Casimir eigenvalues calculated in \cite{Cohen} and those obtained by our formula are shown.\begin{table}[ht]
\caption{Casimir Eigenvalues}     \label{tab:V1}
\small
{
\begin{tabular}{|c|c|c|c|c|c|c|c|c|}
\hline  
&$a$&$\gamma(H)=4+6a$&$\gamma(C)=3+3a$&$\gamma(G)=4+8a$&$t$&$C_{2,0}=(6+4t)/t$&$C_{1,1}=3(t-1)/t$&$C_{1,2}=(8+4t)/t$\\
\hline  
$A_1$ &  $1/2$     & $7$ & $9/2$ & $8$ &  $2$     & $7$ & $9/2$ & $8$  \\
\hline
$A_2$ &   $1/3$    & $6$ & $4$ & $20/3$ &  $3$    & $6$ & $4$ & $20/3$ \\
\hline
$G_2$ & $ 1/4$    & $11/2$ & $15/4$ & $8$ & $4$    & $11/2$ & $15/4$ & $8$ \\
\hline
$D_4$ &   $1/6$    & $5$ &  $7/2$ & $16/3$ &  $6$    & $5$ & $7/2$ & $16/3$\\
\hline
$F_4$ & $1/9$    & $14/3$ & $10/3$ & $14/3$ & $9$    & $14/3$ & $10/3$ & $14/3$ \\
\hline
$E_6$ &  $1/12$    & $9/2$ & $13/4$ & $14/3$ & $12$    & $9/2$ & $13/4$ & $14/3$\\
\hline
$E_7$ & $1/18$    & $13/3$ &  $19/6$ & $40/9$ & $18$    & $13/3$ & $19/6$ & $40/9$  \\
\hline
$E_8$ & $1/30$    & $21/5$& $31/10$ & $64/15$ & $30$    & $21/5$ & $31/10$ & $64/15$\\
\hline  
\end{tabular} 
}
\end{table}

Thus, we see that the Casimir eigenvalues coincide.

  \section{Non-zero Universal Values of Casimir on Zero Representations} 
  
In the recent work of M.Avetisyan and R.Mkrtchyan (\cite{X2k}) a universal formula for dimensions of the $k$-th Cartan power of 
the $X_2$ representation has been obtained. A notable quality of the $X_2(k,\alpha,\beta,\gamma)$ formula is that for the
parameters, corresponding to the $C_N$ algebra it gives $0$ for any $k\geq2$, while we see in the Section 2 that
the Casimir eigenvalues on those irreps are not 0.
 \newline
 A similar situation regarding $A_2$ algebra takes place. The universal decomposition of the symmetric square of the adjoint representation 
 writes as follows:
 $$S^2g=\mathbbm{1}+Y_2(\alpha)+Y_2(\beta)+Y_2(\gamma)$$
 The $Y_2(\beta)$ for $A_2$ is $0$, whilst the Casimir eigenvalue on the same representation is
 $4t-2\beta$. 
 At first glance
it seems natural to expect, that the Casimir eigenvalues on that representations should be equal to 0, while we see,
that they are not.
If one thinks deeper, it is easy to understand, that the Casimir eigenvalue does not have to be equal to 0 on a zero-dimensional
representation.
Indeed, for the points close to the (-2,2,3) on the Vogel plane the Casimir operator acting on the symmetric 
square of the adjoint representation of $A_2$ has three eigenvalues, so in an appropriate basis it has a block-diagonal form.
At (-2,2,3) point all that happens is that $Y_2(\gamma)$ becomes zero for that particular combination of parameters, and the 
corresponding block of the Casimir operator acts on a zero-dimentional vector subspace. Thus we do not see anything that dictates 
that block to be a zero-matrice at that particular point. 
\newline
After the discussion of this situation one concludes, that the universal description sheds a light on the fact, that
it is not just only reasonable, but turns out to be necessary to consider some non-zero eigenvalues of Casimir operators
on non-existing, i.e. zero-dimensional representations. 
Thus, it seems natural to believe, that the universal formulae "take care" of the "invisibility" of that sort of Casimirs.
In other words, we expect that in the universal formulae the Casimir eigenvalues appear in the product with 
the universal dimensions, or, more generally, with expressions, which are necessarily zero, if
the dimension is zero.
\newline
 In support of this idea we bring a formula, presented by Deligne in
 \cite{LaSerie}:
 $$Tr(C_2, [R]V)=\frac{1}{n!}\sum_{\sigma}\chi(\sigma)m(\sigma)(dimV)^{n(\sigma)-1}Tr(C_2, V)$$
 where $V$ is a representation of the algebra, $R$ is a representation of the $S_n$ group, 
 $[R](V):=Hom_{S_n}(R, \otimes^nV)$, $\sigma$ is an element of $S_n$,
  $\chi(\sigma)$ is the character on that element, $m(\sigma)$ is the sum of the squares of the lengths of cycles 
  of $\sigma$, $n(\sigma)$ is the number of cycles of $\sigma$.
\newline
For the symmetric square of the adjoint representation, we rewrite this formula explicitly:
$$1\cdot C_2(\mathbbm{1})+ \text{dim}Y_2(\alpha)C_2(Y_2(\alpha))+ \text{dim}Y_2(\beta)C_2(Y_2(\beta))+ \text{dim}Y_2(\gamma)C_2(Y_2(\gamma))=
(2+ \text{dim}g)\cdot  \text{dim}g C_2(g),$$
where $g$ is the adjoint representation.

Substituting the corresponding universal formulae, one can check, that for $A_2$ algebra this formula is true.

\newpage

\section{Conformity With $sp(-2n)=so(2n)$ Duality}
In (\cite{MV}) R.Mkrtchyan and A.Veselov  have discussed the duality of higher-order Casimir operators for $SO(2n)$ and $Sp(2n)$ groups. Using the Perelomov and Popov 
(\cite{PP}) formula for the generating function for the Casimir spectra and parametrizing the 
Young diagrams in a different way (\cite{MV}), they have explicitly shown the $C_{Sp(2n)}(\lambda,z)=-C_{SO(-2n)}(\lambda',-z)$ duality for 
rectangular Young diagrams.
\newline
Here we write the expressions for the corresponding eigenvalues of the second Casimir operator
($C_2$) for $so(2n)$ and $sp(2n)$ algebras, in the $A, B$ parametrization, used in \cite{MV}. 
\subsection*{$so(2n)$}
For $so(2n)$ the Casimir spectra writes as follows

\begin{align*}
C_{so(2n)}(z,A,B)= \sum_{p=0}^{ \infty}C_{p_{so(2n)}} z^p=
\frac{(1- z n) (2-z (4 n-3))}{z (1-z (n-1)) (2-z (4 n-2))} \times \\
\prod _{i=0}^k \frac{1-z (-A_{k-i}+B_i+2 n-1)}{1-z (A_{k-i}-B_i)} \times
\prod _{i=1}^k \frac{1-z (A_{-i+k+1}-B_i)}{1-z (-A_{-i+k+1}+B_i+2 n-1)}
\end{align*}

After a proper expansion of $C_{so(2n)}(z,A,B)$ into series in the vicinity of the $z_0=0$ point, one can check, that the coefficient of
$z^2$, i.e. $C_{2_{so(2n)}}$ can be expressed as follows:

\begin{align*}
C_{2_{so(2n)}}(A,&B)=  \sum _{i=1}^k \big(4 n A_i (B_{-i+k+1}-B_{k-i})+2 A_i^2 (B_{k-i}-B_{-i+k+1})+ \\
 &+ 2 A_i (B_{k-i}-B_{-i+k+1})+2 B_i^2 (A_{-i+k+1}-A_{k-i}) \big)-4 n A_0 B_k+\\
 &+A_0^2 (2 B_k+4 B_0)+2 A_0 (B_k-B_0)-B_0^2 (2 A_k+4 A_0)-\\
 &-n (A_0-B_0)+2 n \left(A_0^2+B_0^2\right)+2 \left(B_0^3-A_0^3\right)+1/2(A_0-B_0).
\end{align*}

\subsection*{$sp(2n)$}
The Casimir spectra for this case is
\begin{align*}
C_{sp(2n)}(z, &A, B)= \sum_{p=0}^{\infty} C_{p_{sp(2n)}} z^p=\frac{(1- z n) (2-z (4 n+3))}{z (1-z (n+1)) (2-z (4 n+2))} \times  \\
& \prod _{i=0}^k \frac{1-z (B_{k-i}-A_i+2 n+1)}{1-z (-B_{k-i}+A_i)} \times \prod _{i=1}^k \frac{1-z (-B_{-i+k+1}+A_i)}{1-z (B_{-i+k+1}-A_i+2 n+1)}
\end{align*}
And for $C_{2sp(2n)}$ one has
\begin{align*}
C_{2_{sp(2n)}}(A,&B) =  -\sum _{i=1}^k \big(-4 n B_i (A _{-i+k+1}-A _{k-i})+2 A_i^2 (B _{-i+k+1}-B_{k-i})+\\
& 2 B_i^2 (A_{k-i}-A_{-i+k+1})++2 B_i (A_{k-i}-A_{-i+k+1})\big)-4n B_0 A_k+\\
&+A_0^2 (2 B_k+4B_0)-2 B_0 (A_k-A_0)-B_0^2 (2 A_k+4A_0)-\\
&-n (B_0-A_0)+2 n \left(A_0^2+B_0^2\right)+1/2(A_0-B_0)-2 \left(A_0^3-B_0^3\right).
\end{align*}

Therefore, we have obtained formulae for second Casimir eigenvalues on irreps of  $sp(2n)$ and $so(2n)$ algebras,
corresponding to any Young diagram (any $(A,B)$ set).
\newline
It can be checked, that
$$C_{2_{so(2n)}}(A,B)=-C_{2_{sp(-2n)}}(B,A)$$
i.e. the Casimir duality for the second Casimir holds for any Young diagram (for any $A,B$ set).
In particular, for $X_2$ one has the values, shown in the Table 4.
It can be observed, that $C_{2_{so(2n)}}=2C_{2_{sp(2n)}}=1/2C_{2_{so(2n)}}(A,B)$, which indicates the difference of the definition of the
Killing form in \cite{MV} \footnote{in \cite{MV} the Killing form is defined as $Tr(\hat{X^a}, \hat{X^b})$ in the fundamental representation, while our normalization (so called Cartan-Killing normalization)
corresponds to the Killing form, defined as $Tr(ad\hat{X^a}, ad\hat{X^b})$, i.e. in the adjoint representation. }.

\begin{table}
\caption{Comparison} 

\begin{tabular}{|c|c|c|c|c|}
\hline  
$Algebra$ & $ Diagram$ & $A,B$ & $C_2(A,B)$ & $C_2$ \\
\hline  
$so(2n)$ &  (2,1,1) & $A_1=B_1=1,A_2=3,B_2=2$  & $16n-16$ & $8n-8$ \\
\hline
$sp(2n)$ &   (3,1) & $A_1=B_1=1,A_2=2,B_2=3$   & $16n+16$ & $4n+4$ \\
\hline
\end{tabular}
\end{table}
 
 In \cite{X2k} it has been shown, that when permuting the Vogel parameters
 corresponding to the $so(2n)$ algebra in this way: $(\alpha, \beta, \gamma)\to (\beta, \alpha, \gamma)$, the $X_2(k)$ formula
 gives dimensions for some representations of the $sp(2n)$ algebra. More precisely, that permutation specifies a correspondence
 between $\lambda^{so(2n)}=k(\omega_1+\omega_3)$ and $\lambda^{sp(2n)}=2\omega_k+\omega_{2k}$ representations.
 One can notice, that the Young diagrams, associated with these representations are conjugate with each other. Indeed, in $A,B$
 parametrization the associated sets are $$\lambda^{so(2n)} \leftrightarrow {A_0=B_0=0, A_1=1, B_1=k, A_2=3, B_2=2k},$$
 $$\lambda^{sp(2n)} \leftrightarrow {A_0=B_0=0, A_1=k, B_1=1, A_2=2k, B_2=3}.$$
 Therefore, it is reasonable to check the Casimir duality for these representations.
 Substituting the corresponding $(A,B)$ sets into the expressions for $C_2(A,B)$ written above, one gets
 $$C_{2_{so(2n)}}(A,B)=12k^2+k(16n-28),$$
 $$C_{2_{sp(2n)}}(B,A)=-12k^2+k(16n+28)=-(12k^2+k(16(-n)-28)=-C_{2_{so(2n)}}(A,B).$$
 So, the Casimir duality holds for representations, associated with the  \newline
 $X_2(k,-2,4,2n-4) \leftrightarrow X_2(k,4,-2,2n-4)$ transformation
 of the $X_2(k,\alpha,\beta,\gamma)$ universal formula \cite{X2k}.\newline
 For the same representations in the Cartan-Killing normalization we have $$C_{2_{so(2n)}}=6k^2+k(8n-14),$$  $$C_{2_{sp(2n)}}=-3k^2+k(4n+7),$$
 i.e. $$C_{2_{so(2n)}}(\lambda)=-2C_{2_{sp(-2n)}}(\lambda'),$$ as expected.
 \newpage
 
\section{Acknowledgements}
I am grateful to Professor R. Mkrtchyan for his valuable guidance and productive discussions.
\newline
This work was fulfilled within the Regional Doctoral Program on Theoretical and Experimental Particle Physics sponsored by 
VolkswagenStiftung and was partially supported by the Science Committee of the Ministry of Science and Education of the Republic of Armenia
 under contract 18T-1C229.


\begin{thebibliography}{66}
\bibitem{V0}
 P. Vogel,  The Universal Lie algebra. Preprint (1999),
 https://webusers.imj-prg.fr/\~{}pierre.vogel/grenoble-99b.pdf

\bibitem{V}
P.Vogel,   Algebraic structures on modules of diagrams. Preprint (1995), 
www.math.jussieu.fr/\~{}vogel/diagrams.pdf, J. Pure Appl. Algebra {\bf 215} (2011),
no. 6, 1292-1339. 

\bibitem{LM1} 
J.M. Landsberg and  L.Manivel,  A universal dimension formula for complex simple
Lie algebras. Adv. Math. {\bf 201} (2006), 379-407

\bibitem{MSV}
R.L. Mkrtchyan, A.N. Sergeev and A.P. Veselov,  
Casimir eigenvalues for universal Lie algebra, arXiv:1105.0115,  Journ.
Math.Phys. 53, 102106 (2012).

\bibitem{MV}
R.Mkrtchyan, A.Veselov, On duality and negative dimensions in the theory of Lie groups and symmetric spaces
Journal of Mathematical Physics 52, 083514 (2011)

\bibitem{M13}
   R.L.Mkrtchyan,  Nonperturbative universal Chern-Simons theory. 
   JHEP09(2013)54, arXiv:1302.1507.
   
 \bibitem{KhM16-1}
H.M.Khudaverdian, R.L.Mkrtchyan,  Universal volume of groups 
 and anomaly of Vogel's symmetry, arXiv:1602.00337, Letters in Mathematical
 Physics, (2017) 107(8), 1491-1514, DOI 10.1007/s11005-017-0949-8,   
 
  \bibitem{M17}
R.L.Mkrtchyan, Partition function of Chern-Simons theory as renormalized
q-dimension, arXiv:1709.03261, Journal of Geometry and Physics, Volume 129, July
2018, Pages 186-191   


\bibitem{W15}
B.W.Westbury, Extending and quantising the Vogel plane, arXiv:1510.08307.

\bibitem{MMM}
A. Mironov, R. Mkrtchyan and A. Morozov,  On universal knot polynomials,
arxiv:1510.05884 , JHEP02(2016)078

\bibitem{MM}
A.Mironov and A.Morozov, Universal Racah matrices and adjoint knot polynomials.
I. Arborescent knots, arXiv:1511.09077, Physics Letters B755 (2016) 47-57.

\bibitem{M16QD}
R.L.Mkrtchyan, On Universal Quantum Dimensions, arxiv:1610.09910, Nuclear Physics
B921,  2017, pp. 236-249, 

\bibitem{RJ} M. Rosso and V. F. R. Jones, J. Knot Theory Ramifications, \textbf{2} (1993) 97-112

\bibitem{LieART}
R.Feger, T.W.Kephart, LieART Mathematica Application for Lie Algebras and Representation Theory, ePrint, arXiv:1206.6379v2

\bibitem{Cohen}
A. M.Cohen and R. de Man,  Computational evidence for Deligne’s conjecture regarding exceptional Lie groups, 
Comptes Rendus de l’Academie des Sciences, Serie 1, Mathematique, (1996) 322(5), 427-432

 \bibitem{X2k}
M.Y.Avetisyan, R.L. Mkrtchyan,  $X_2$ series of universal quantum dimensions, ePrint, arXiv:1812.07914
   

\bibitem{LaSerie} 
P.Deligne, La serie exceptionnelle des groupes de Lie, C. R. Acad. Sci. Paris,
Serie I, 322 (1996), 321-326.

\bibitem{MV}
   R.L. Mkrtchyan  and A.P.Veselov,  Universality in Chern-Simons theory. JHEP08
   (2012) 153, arXiv:1203.0766.
   
   
\bibitem{PP}
A. M. Perelomov and V. S. Popov, Math. USSR, Izv. 2(6) , 1313 (1968)



\end{thebibliography}
\end{document}